\documentclass[twocolumn,showpacs,showkeys,amsmath,amssymb]{revtex4}
\usepackage{times}
\usepackage{graphicx}

\newcommand{\beq}{\begin{equation}}
\newcommand{\eeq}{\end{equation}}

\def\ba{\begin{eqnarray}}
\def\ea{\end{eqnarray}}

\def\bfr{{\bf r}}
\def\bfv{{\bf v}}
\def\bfx{{\bf x}}
\def\bfg{{\bf g}}

\def\L{\Lambda}
\def\Om{\Omega}
\def\d{\delta}

\def\gs{\mathrel{\lower0.6ex\hbox{$\buildrel {\textstyle >}\over{\scriptstyle \sim}$}}}
\def\ls{\mathrel{\lower0.6ex\hbox{$\buildrel {\textstyle <}\over{\scriptstyle \sim}$}}}

\begin{document}

\title{Evolution of gravitational orbits in the expanding universe}

\author{Mauro Sereno}
\email{sereno@physik.unizh.ch}

\author{Philippe Jetzer}
\email{jetzer@physik.unizh.ch}

\affiliation{Institut f\"{u}r Theoretische Physik, Universit\"{a}t Z\"{u}rich,
Winterthurerstrasse 190, CH-8057 Z\"{u}rich, Switzerland}


\begin{abstract}
The gravitational action of the smooth energy-matter components filling in the universe can affect the orbit of a planetary system. Changes are related to the acceleration of the cosmological scale size $R$. In a universe with significant dark matter, a gravitational system expands or contracts according to the amount and equation of state of the dark energy. At present time, the Solar system, according to the $\L$CDM scenario emerging from observational cosmology, should be expanding if we consider only the effect of the cosmological background. Its fate is determined by the equation of state of the dark energy alone. The mean motion and periastron precession of a planet are directly sensitive to $\ddot{R}/R$, whereas variations with time in the semi-major axis and eccentricity are related to its time variation. Actual bounds on the cosmological deceleration parameters $q_0$ from accurate astrometric data of perihelion precession and changes in the third Kepler's law in the Solar system fall short of ten orders of magnitude with respect to estimates from observational cosmology. Future radio-ranging measurements of outer planets could improve actual bounds by five order of magnitude.
\end{abstract}


\maketitle

\section{Introduction}

The local effect on particle dynamics from the global expansion of the universe is a classical topic in general relativity. The picture is pretty clear for a test particle moving in the homogeneous and isotropic Robertson-Walker (RW) metric. Analytical methods are mainly based on solving the geodesic equations \cite[and references therein]{gr+el06} and have been used to examine whether particles removed from the Hubble flow asymptotically rejoins it. Different definitions of what this means can lead to apparent disagreement \cite{bar+al06}. A general result for a test particle in a RW metric is that its peculiar velocity with respect to the Hubble flow declines in magnitude as the scale factor of the universe, $R$, expands, alike to any photon energy \cite{pea01}. This characteristic does not assure that all of the features of a generic trajectory in the expanding space will approach those of a trajectory in the Hubble flow. As a matter of fact, many characteristics depend on the acceleration, or deceleration, of the universe rather than just its expansion. The strong requirement that the proper distance between the test particle and a reference asymptotic particle approaches zero holds only in an accelerated universe. In an Einstein-de Sitter universe, a particle initially at rest with respect to the origin falls towards the origin, passes through it and asymptotically moves on the opposite side of the sky with a null peculiar velocity \cite{pea01, whi04}. Anyhow, the spatial 'peculiar' distance between the test particle and the background particle with the same asymptotic speed is unbounded and increases with time \cite{whi04}. Even for an under-dense universe, the particle always keeps a distance from its corresponding background place \cite{whi04}. A cosmological constant can change this picture. Due to the repulsive effect of the vacuum energy, the peculiar distance decreases asymptotically. Nevertheless, the effect does not arise from the expansion of the universe but stems from the gravitational repulsion of the new ingredient in the cosmic energy budget \cite{pea01, whi04}.

Matter inhomogeneities complicate the picture. In the Einstein-Strauss model, the strict vacuum spherically symmetric Schwarzschild solution is truncated at a finite radius and matched to a cosmological model. The Swiss cheese universe generalizes this solution, prescribing that the mass within non overlapping spherical cavities in an otherwise homogeneous and isotropic expanding universe is compressed to a point mass. During the expansion the only change of any cell is an increase of the truncation radius but, within the cavity, the gravitational field remains exact Schwarzschild and the behavior of the rest of the universe is irrelevant. This could be a compelling argument on the (absence of) cosmological influence on gravitationally bound systems, its obvious shortcoming being that it does not consider the gravitational effect of the (expanding) energy-matter fluid wherein the test particles move. In analogy with the RW metric, it should be not the expansion of the universe that can cause gravitational bound systems to separate or evolve. The effect, if any, should stem from the gravitational action of the homogeneous background in which the system is embedded. How the expansion of the universe affects the dynamics of local systems has been analyzed mainly considering the two-body problem in a cosmological background \cite{mcv33,di+pe64,no+pe71,and95,coo+al98}. Some apparent differing results could spring from the different coordinate systems adopted in different studies. The equation of motion are not coordinate invariant and the form of any correction due to the cosmic background is dependent upon the frame employed \cite{coo+al98}. In fact, many apparent discrepancies disappear if the different approaches are performed coherently \cite {ca+gi06}.
 
In a series of previous papers, we discussed the effect on a local bound system of the cosmological constant \cite{je+se06,se+je06a} and of diffuse dark matter \cite{se+je06b}. In this paper we investigate the evolution of a planet orbit around a massive star embedded in an homogeneous and isotropic universe. In Section~\ref{sec:eul}, we use some arguments based on Newtonian dynamics to re-derive the relevant equations of motion for a test particle in a RW universe perturbed by a weak local inhomogeneity. Section~\ref{sec:evo} is devoted to study the effect of dark matter and dark energy on the orbital radius evolution of a bound system. In Section~\ref{sec:pla}, the Lagrange's planetary equations are discussed and the time variation of the orbital elements is derived. Section~\ref{sec:con} is devoted to some final comments.

\section{Equation of motion}
\label{sec:eul}

The essential physics of the motion of a test particle in the weak gravitational field of a local inhomogeneity perturbing an otherwise homogeneous and isotropic expanding universe can be extracted following the Newtonian approach as usually performed for the dynamics of linear perturbations \cite{pea99}. The background is suitably described by the RW metric. The well known equations for the evolution of the scale factor $R$ are
\ba
\dot{R}^2 &  =&  \frac{8\pi G}{3}\rho R^2-kc^2,   \label{fri1} \\
\ddot{R}  &  =& -\frac{4\pi G}{3} R (\rho +3 \frac{p}{c^2}), \label{fri2}
\ea
where dot stands for time derivative, $G$ is the Newton's gravitational constant, $c$ the speed of light in the vacuum, $\rho$ and $p$ the density and pressure of the total cosmic fluid (inclusive of any dark energy component), respectively, and $k$ a curvature constant. Equation~(\ref{fri1}) is known as the Friedmann equation and can be also seen as an equation for the Hubble parameter $H=\dot{R}/R$. The scale factor $R$ can be expressed in terms of the redshift $z$ as $R=1/(1+z)$, where we have normalized $R=1$ at the present time.

Let us introduce Eulerian coordinates $\bfx$. The corresponding comoving spatial coordinates are $\bfr = \bfx (t)/ R(t)$. The time derivative of $\bfx$ is the sum of the Hubble expansion plus the peculiar velocity, $\dot{\bfx} = H \bfx +\delta {\bfv}$, where the peculiar velocity $\delta \bfv$ can be expressed as the time derivative of the comoving coordinate,  $\delta \bfv = R~\dot{\bfr}$.  Differentiating $\bfx$ twice gives
\beq
\label{eul1}
\ddot{\bfx} = \dot{\delta \bfv} +H \delta {\bfv} +\frac{\ddot{R}}{R} \bfx .
\eeq
The Newtonian limit of the equation of motion for a test particle in the RW metric is obtained from Eq.~(\ref{eul1}) for zero peculiar velocity and zero peculiar acceleration, $\ddot{\bfx}= (\ddot{R}/R) \bfx$.

The perturbed equation of motion follows from the Eulerian equation of motion
\beq
\label{eul2}
\ddot{\bfx} = \bfg_0+ \bfg  ,
\eeq
with $\bfg_0$ being the unperturbed gravitational acceleration that acts on a particle in a homogeneous universe and $\bfg$ the peculiar gravitational acceleration. We remind that $\bfg_0$ can not be derived from Newtonian gravity alone and must be assumed to be given \cite{pea99}. 

Identifying the right hand sides of Eqs.~(\ref{eul1},~\ref{eul2})  in the case of zero peculiar terms, we get the unperturbed equation, $(\ddot{R}/R)\bfx = \bfg_0$. Then, subtracting such an unperturbed equation from Eq.~(\ref{eul2}) side by side, we get
\beq
\label{eul6}
\ddot{\bfx} - \frac{\ddot{R}}{R}\bfx = \bfg .
\eeq
The cosmological influence on the local dynamics of a bound system appears through the acceleration of the scale factor and not through its first derivative. Equation~(\ref{eul6}) has been already derived several times in literature with ${\bf g}$ specified to the case of a point mass, ${\bf g} = - G M \hat{\bfx}/x^2$ \cite{coo+al98,ca+gi06}. \citet{coo+al98} derived the geodesic equations for a RW metric in the local inertial frame using Fermi normal coordinates and then added linearly the peculiar Newtonian acceleration due to a massive body. \citet{ca+gi06} reviewed different approaches leading to Eq.~(\ref{eul6}). In particular they considered both the motion of a test particle in the McVittie spacetime, already considered in \citep{pac63}, and the full relativistic treatment for electromagnetically-bounded systems, drawing on some arguments first discussed in \cite{di+pe64}.

With the aid of Eq.~(\ref{eul1}), equation~(\ref{eul6}) can be rewritten in terms of the peculiar velocity as
\beq
\label{eul3}
\dot{\delta \bfv} +H \delta {\bfv} = \bfg .
\eeq
The solution of equation~(\ref{eul3}) is
\beq
\label{eul5}
\delta {\bfv}(t) = \frac{1}{R(t)} \left( A + \int^t R(t') \bfg (t') d t' \right) ,
\eeq
with $A$ a constant of integration. In the unperturbed case, peculiar velocities of non-relativistic objects declines as $1/R(t)$, just as any photon momenta. Of course, in presence of a local inhomogeneity in the matter distribution, $\delta v$ does not tend to zero anymore. Dot in the previous equations stands for the full convective time derivative, which when dealing with perturbed quantities can be replaced by $d/dt\equiv~\partial~/~\partial~t~+~\bfv_0~\cdot~\nabla$. For a point-mass perturbation in a uniform background, $d(\delta \rho_0/\rho_0)/dt= 0$, and the perturbed linearized equation for conservation of energy takes the form $ \nabla \dot{\delta\bfv} = 0$. Then, the convective derivative can be identified with the partial time derivative, $d \delta {\bfv}/dt  =\partial \delta {\bfv}/\partial t$.

In the following sections, Equation~(\ref{eul6}) will be specified  in the case of a central massive body, ${\bf g} = - G M \hat{\bfx}/x^2$, and will be the starting point for our considerations.

\section{Evolution of the orbital radius}
\label{sec:evo}

\begin{figure}
\resizebox{\hsize}{!}{\includegraphics{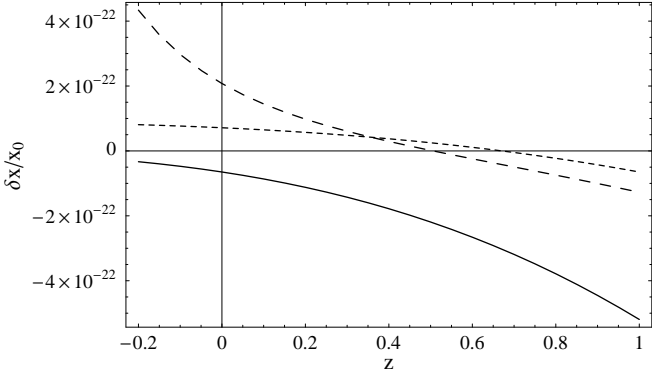}} 
\caption{Orbital radius evolution for an Earth-like planet (orbital period $\sim 1$~year). The variation is put to zero at the transition redshift $z_\mathrm{tr}$ from a decelerated to an accelerated expansion. We consider flat models with $h=0.7$. The full line is for an Einstein-de Sitter model $\Om_\mathrm{M0}=1$ (the acceleration gets asymptotically to zero at $z_\mathrm{tr} \rightarrow -1$), the dashed line for a $\L$CDM model with $\Om_\mathrm{M0}=0.3$ and $w_\mathrm{X0}=-1$ ($z_\mathrm{tr} \simeq 0.67$), the long-dashed line for $\Om_\mathrm{M0}=0.3$ and phantom energy with $w_\mathrm{X0}=-2$ ($z_\mathrm{tr} \simeq 0.51$).}
 \label{deltar_z}
\end{figure}

\begin{figure}
\resizebox{\hsize}{!}{\includegraphics{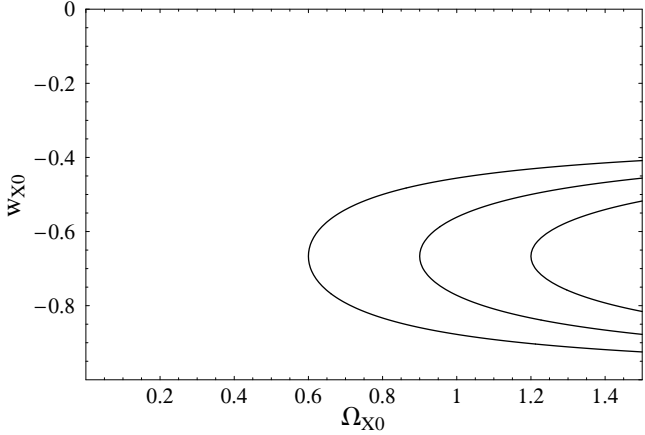}} 
\caption{The curves represent the loci of points in the space of dark energy parameters, i.e. the $\Om_\mathrm{X0}-w_\mathrm{X0}$ plane, for which the radius of a gravitationally bound system does not vary at the present time. Curves from the left to the right are for $\Om_\mathrm{M0}=0.2,0.3$ and 0.4, respectively. The region on the left side of each curve corresponds to positive expansion for the given $\Om_\mathrm{M0}$.}
 \label{OmegaX_wX_exp}
\end{figure}

\begin{figure}
\resizebox{\hsize}{!}{\includegraphics{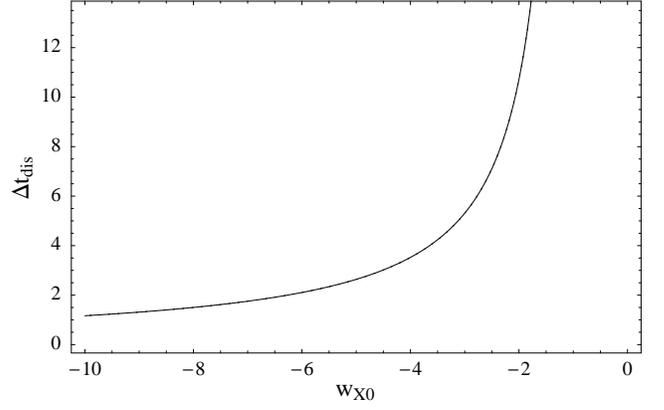}} 
\caption{ Time (in units of $10^9/h$~years) needed for the dissociation of an Earth-like system ($P\sim 1$~year) in a phantom cosmology as a function of the equation of state of the dark energy. We have considered a flat model with $\Om_\mathrm{M0}=0.3$.}
 \label{wx_deltat}
\end{figure}

The motion of a test particle in the expanding universe is a typical multiple time scale problem in which the natural period of the unperturbed orbit is much smaller than the time scale over which the perturbation changes, i.e. the age of the universe. Equation~(\ref{eul6}) can be conveniently rewritten in cylindrical coordinates $\{ x,\varphi\}$ \citep{coo+al98,ne+pe04}. For a test particle around a massive central object $M$, it is
\ba
\ddot{x} - x\dot{\varphi}^2 + \frac{G M}{x^2} -\frac{\ddot{R}}{R} x & = & 0 \label{eul7} \\
 x^2 \dot{\varphi}  & = & L_0 \label{eul8} .
\ea
with $L_0$ the conserved angular momentum per unit mass. Since the perturbation is purely radial, the motion is confined to the unperturbed Keplerian orbital plane. An approximate solution to the orbital motion can be found if we consider the perturbations of the orbital coordinates in the case of circular orbits,
\ba
x(t)        & = & x_0   + \d x(t),  \label{eul9}  \\
\varphi (t) & = & n_0 t + \d \varphi(t) \label{eul10},
\ea
with $x_0$ and $n_0=\sqrt{G M /x_0^3}$ being the unperturbed orbital radius and mean motion, respectively. We assume that the time dependence of the perturbations in Eqs.~(\ref{eul9},~\ref{eul10}) is due to the time variation of the perturbing function. Substituting in Eqs.~(\ref{eul7},~\ref{eul8}) yields
\beq
\ddot{\d x} + n_0^2 \d x  -\frac{\ddot{R}}{R} x_0  =  0 . \label{eul11}
\eeq
As far as we consider secular variations, the $\ddot{\d x}$ term in the above equation can be neglected since the unperturbed orbital period is much smaller than the universe age \cite{coo+al98}. Then,
\beq
\label{eul12}
x(t) \simeq x_0 \left( 1+ \frac{1}{n_0^2} \frac{\ddot{R}}{R} (t) \right) .
\eeq
This can be also seen by writing down the equation for the centripetal acceleration in a circular motion,
\beq
\label{eul13}
\dot{\varphi}^2 x = \frac{G M}{x^2} -\frac{\ddot{R}}{R} x , 
\eeq
where $\dot{\varphi}$ is the angular frequency that we can assume as constant during a planet orbit. Remembering that $\dot{\varphi} x^2 = L_0 = n_0 x_0^2$, we get
\beq
\label{eul14}
1= \frac{x}{x_0} -\frac{1}{n_0^2}\frac{\ddot{R}}{R} \left(  \frac{x}{x_0} \right)^4 .
\eeq
Inserting an expansion for $x$ such as in Eq.~(\ref{eul9}) in the above equation, we re-obtain the result for the secular evolution, Eq.~(\ref{eul12}). The unperturbed orbital radius can be fixed to its value at the time when the perturbation cancels out, i.e. when the expansion of the universe changes from a decelerated expansion to an accelerated one. 

In what follows, we will consider a model of universe filled in with pressureless dark-matter, $\Om_\mathrm{M0}$, and dark energy $\Om_\mathrm{X0}$ with a constant equation of state, $w_\mathrm{X0}$. Then the field equations for the scale factor, Eqs.~(\ref{fri1},~\ref{fri2}), take the form
\ba
\left( \frac{H}{H_0}\right)^2  & = & \Om_\mathrm{M0}R^{-3}+\Om_\mathrm{X0}R^{-3(1+w_\mathrm{X0})} + \Om_{K0}R^{-2} , \label{fri3} \\
\frac{\ddot{R}}{R} & = & -\frac{H_0^2}{2}  \label{fri4} \\
& \times & \left\{ \Om_\mathrm{M0} R^{-3}  + (1+3 w_\mathrm{X0} ) \Om_\mathrm{X0} R^{-3(1+w_\mathrm{X0}) } \right\}, \nonumber
\ea
with $\Om_{K0} = 1- \Om_\mathrm{M0} -\Om_\mathrm{X0} $. The Hubble constant can be written in convenient units as $H_0= h \times 100$~km~s$^{-1}$Mpc$^{-1}$. For our reference model, we will mean a flat $\L$CDM model with $\Om_\mathrm{M0}=0.3$ and dark energy in the form of a cosmological constant, $w_\mathrm{X0}=-1$.

At the transition redshift
\beq
z_\mathrm{tr} =\left\{ -(1+3 w_\mathrm{X0} )\frac{\Om_\mathrm{X0}}{\Om_\mathrm{M0}} \right\}^{-1/(3 w_\mathrm{X0}) } -1,
\eeq
the acceleration cancels out, $\ddot{R}(z_\mathrm{tr})=0$. Then, we will fix the orbital parameters to their unperturbed values at $z=z_\mathrm{tr}$. In Fig.~\ref{deltar_z} we plot the time evolution of the orbital radius for an Earth-like planet (orbital period $P \sim 1$~year). Relative variations around our present time are really small. 

The behavior of an orbit depends on which energy-matter component of the universe stands out. Dark matter has an attractive effect and its density falls with time for an expanding universe. So, for a CDM model without dark energy, the orbital radius expands as far as the scale factor expands. On the other hand, dark energy has a repulsive gravitational effect that can balance the dark matter action. Let us consider the condition for the expansion (contraction) of a gravitationally bound system, $d (\ddot{R}/R)/ dt > (<) 0$. The rate of fractional change of the orbital radius can be expressed as a function of redshift and cosmological parameters as 
\ba
\frac{1}{x} \frac{dx }{dt} & = & \frac{3}{2} \left( \frac{H_0}{n_0}\right)^2 H(z) R^{-3} \\
& \times & \left\{ (1+ w_{\text{X0}})( 1+3 w_{\text{X0}}) \Omega_{\text{X0}} R^{-3 w_\text{X0} }+ \Omega_{\text{M0}} \right\} . \nonumber
\ea
The fate of the bound system depends on the equation of state of the dark energy. In fact, if any dark energy is present, it will determine the expansion of the universe at future times, with no regard to the matter content. For either $w_\mathrm{X0}> -1/3$, which means a decelerated expansion of the scale factor of the universe, or $w_\mathrm{X0} <-1$, the orbital radius will expand with no regard to the total amount of dark matter and dark energy, see Fig.~\ref{OmegaX_wX_exp}. For $w_\mathrm{X0} > -1/3$, the total source of gravitational attraction $\rho+3p$ is positive and the dark energy is not repulsive anymore. The perturbation to the radius, which is $\propto \ddot{R}/R$, stays negative and decreases in absolute value towards zero as the density decreases with time. The rate of change is as well decreasing, $d x/d t \rightarrow 0$ for $R \rightarrow \infty$. For $-1 <w_\mathrm{X0} < -1/3 $, dark energy is effectively repulsive ($\ddot{R}/R >0$ at late times). Its density and its repulsive effect drop with time. Then, the perturbation is positive and the orbital radius evolves towards its unperturbed value when the scale factor $R$ diverges. In the case of dark energy in the form of a cosmological constant, $w_\mathrm{X0}=-1$, the universe is destined to a de Sitter inflationary expansion during which both the orbital radius and its positive perturbation remain asymptotically constant. 

In phantom cosmologies, $w_\mathrm{X0}<-1$, the dark energy density grows with time and will eventually diverge in a finite time. The acceleration of the scale factor increases towards a singularity which is characterized by the divergences of the scale factor, the Hubble parameter $H$ and its derivative $\dot{H}$ and the scalar curvature \cite{cal+al03}. The repulsive gravitational action of the phantom energy will overcome the normal binding force of structures which will get dissociated \cite{cal+al03,ne+pe04}. As for the case $-1/3 <w_\mathrm{X0} <-1 $, the perturbation is positive at late time, but both the perturbation, $\ddot{R}/R$, and the rate of change of the perturbation, $d(\ddot{R}/R)/dt$, grow with time. The orbital radius becomes larger with time with no regard to the matter content and it will get eventually unbound in a finite time.

While the universe undergoes from a dark matter to a dark energy dominion, the orbit of a gravitationally bound system will move from a contracting to an expanding phase. The inversion redshift is
\beq
z_\mathrm{inv} = - 1+(1 + w_\mathrm{X0} )^{- 1/(3 w_\mathrm{X0})} ( z_\mathrm{tr} + 1 ) .
\eeq
Such an inversion trails the transition from a dark matter to a dark energy dominated universe, $z_\mathrm{tr}$. As we have seen before, the inversion would happen only for dark energy with equation of state $-1 < w_\mathrm{X0} < - 1/3$.

Let us now discuss the situation at the present time. The orbital radius is growing just now if 
\beq
\left. \frac{d}{dz}\left( \frac{\ddot{R}}{R}  \right) \right|_{z=0} < 0 ;
\eeq 
the condition for expansion can be expressed in terms of cosmological parameters as,
\beq
\label{pla9}
\Om_\mathrm{M0} > -(1+w_\mathrm{X0})(1+3 w_\mathrm{X0})\Om_\mathrm{X0} .
\eeq
The equality between the sides of Eq.~(\ref{pla9}) guarantees for a constant orbital radius. We can easily see how, for our reference $\L$CDM model, a planetary system is expected to be expanding. This condition would change for $\Om_\mathrm{M0} \ls 0.2$ and $w_\mathrm{X0} \gs -1$, see Fig.~\ref{OmegaX_wX_exp}.

In the phantom scenario, the orbit radius will eventually diverge. Then, the perturbation approach is not enough to track the complete evolution of the system, but alternative analytical methods can be used \cite{ne+pe04,ca+gi06}. The dynamics of the radial motion of the bound system can be seen as determined by an effective potential. Assuming circular orbits,
\beq
\label{pot1}
V_\mathrm{eff}= - x_0^2 \left\{ \frac{n_0^2}{(x/x_0)}  -  \frac{n_0^2}{2 (x/x_0)^2}  + \frac{1}{2}\frac{\ddot{R}}{R}(t) (x/x_0)^2 \right\} .
\eeq
The equation for the minimum of the potential is the same of Eq.~(\ref{eul14}). The minimum with respect to $x$ at time $t$ gives the approximate orbital radius. The minimum exists only for \cite{ne+pe04}
\beq
\label{pot2}
\frac{1}{n_0^2}\frac{\ddot{R}}{R}(t) \leq \frac{27}{256} ,
\eeq
with the equality picking out the time when the minimum disappears and the system becomes unbound. The dissociation time depends nearly exclusively on the late time behavior of the dark energy density. Then, we can find an approximate solution for the redshift of dissociation by considering $\Om_\mathrm{M0}=0$. We get
\beq
\label{pot3}
z_\mathrm{dis} = -1 +\left\{ \frac{3}{4}\left( \frac{n_0}{H_0}\right)^{2/3} \left[ -2(1+3w_\mathrm{X0}) \Om_\mathrm{X0} \right] ^{-1/3}   \right\}^\frac{1}{1+w_\mathrm{X0} } .
\eeq
The span of time between today and when the system gets dissociated is
\beq
\label{pot4}
\Delta t_\mathrm{dis} = \int_{z_\mathrm{dis}}^0 \frac{d z}{ (1+z)H(z) } .
\eeq
As far as $n_0 \ll H_0$, the dissociation redshift $z_\mathrm{dis}$ will be nearly indistinguishable from $-1$, with no regard to the properties of the dark energy. Then, $\Delta t_\mathrm{dis}$ will be nearly independent from the orbital period of the system and the dissociation will take place nearly at the same time of the future singularity. In Fig.~\ref{wx_deltat}, we plot $\Delta t_\mathrm{dis}$ as a function of the equation of state. We can see that dissociation due to dark energy is not an impending danger.

\section{Element perturbations}
\label{sec:pla}

\begin{figure}
\resizebox{\hsize}{!}{\includegraphics{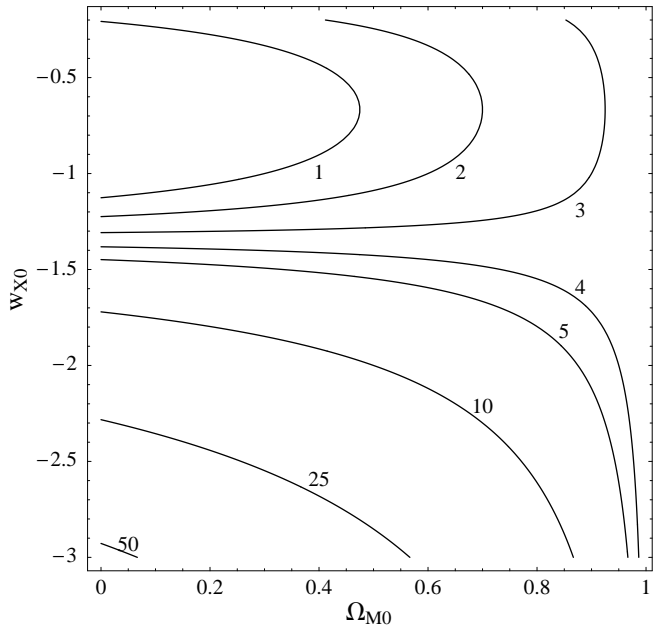}} 
\caption{Rate of variation of the orbital radius in the $\Om_\mathrm{M0}-w_\mathrm{X0}$ plane in units of the rate occurring for a standard flat model with cosmological constant and $\Om_\mathrm{M0}=0.3$. We assume flat models.}
 \label{OmegaM_wX_var}
\end{figure}

A perturbed planet orbit can be described through the orbital elements of its osculating ellipse \citep{dan88}. The Lagrange's planetary equations account for time variation of these elements. For a radial perturbation ${\cal{A}}_r$, we have \citep{dan88}
\ba
\frac{d a}{dt} & = & \frac{2}{n}\frac{e}{\sqrt{1-e^2}} {\cal{A}}_r \sin \varphi  , \label{pla1} \\
\frac{d e}{dt} & = & \frac{1}{n a}\sqrt{1-e^2} {\cal{A}}_r\sin \varphi , \label{pla2} \\
\frac{d \omega}{dt} & = & - \frac{1}{n a}\frac{\sqrt{1-e^2}}{e} {\cal{A}}_r \cos \varphi  \label{pla3} ,
\ea
where $a$ is the semi-major axis, $e$ the eccentricity, $n \equiv \sqrt{G M/a^3}$ the mean motion, $\omega$ the argument of periastron and $\varphi$ is the angle in the orbit plane from the periastron. Since the perturbation is radial, the inclination $i$ and the longitude of the ascending node $\Omega$ do not change. 

In order to reveal the slow secular variation of the orbital elements due to the cosmological expansion, the above relations can be averaged over the unperturbed orbital motion,
\ba
x & = & \frac{a(1-e^2)}{1+e\cos \varphi}, \label{pla4}\\ 
\frac{d \varphi}{d t} & = & \frac{L_0}{ r^2}, \label{pla5} 
\ea
with $L_0=\sqrt{G M a (1-e^2)} $. The average of a generic quantity $f$ over an unperturbed period $P=2\pi/n$ is performed through the integral 
\beq
\left\langle f \right\rangle = \frac{(1-e^2)^{3/2}}{2\pi}\int_{0}^{2\pi} \frac{f(\varphi)d\varphi}{(1+e\cos \varphi)^2}. \label{pla6}
\eeq
The cosmological term $\ddot{R}/R$ can be Taylor expanded for an orbit as
\beq
\frac{\ddot{R} }{R} = \left. \frac{\ddot{R} }{R} \right|_{\varphi =0} + \left. \frac{d }{d \varphi} \left( \frac{\ddot{R}}{R} \right) \right|_{\varphi =0}  \varphi \label{pla7};
\eeq
Averaging over a period, we obtain the rate of variation of the semi-major axis \footnote{In the primitive function of the integral to be performed, there appear terms in a form like $\varphi -2 \tan^{-1}(\tan (\varphi/2))$. The zero order contribution to Eq.~(\ref{pla8}) rises from the fact that the function $\tan^{-1}$ has values between $\pi/2$ and $\pi/2$ whereas the integration range is $0 \leq \varphi \leq 2\pi$.},
\beq
\left\langle \frac{d a}{dt} \right\rangle_t = \frac{a}{n^2}  \left. \frac{d }{dt} \left( \frac{\ddot{R} }{R} \right) \right|_t \left\{ 1+ (1-e^2)\left(\frac{(1-e)^2}{\sqrt{1-e^2}}-1 \right) \right\} . \label{pla8}
\eeq
We turned to the time derivative of $\ddot{R}/R$ thanks to Eq.~(\ref{pla5}). The dependence on the eccentricity is pretty small, with a maximum decrement of $\sim 25\%$ for $e\sim 0.3$ and a null variation from the circular case for very large eccentricity, $e \rightarrow 1$. For a small eccentricity,
\beq
\left\langle \frac{d a}{dt} \right\rangle_t =\frac{a}{n^2}  \left. \frac{d }{dt} \left( \frac{\ddot{R} }{R} \right) \right|_t  \left\{ 1-2e+ {\cal{O}}(e^2) \right\}.
\eeq
For a circular orbit, $e=0$, Eq.~(\ref{pla8}) agrees with the result in Eq.~(\ref{eul12}). The zero order term in the expansion in Eq.~(\ref{pla7}) does not affect $a$, which changes only for a scale factor with a non constant acceleration. In a model of universe dominated by a cosmological constant (de Sitter universe), the orbital radius does not change in time. This result can be also obtained in the framework of the Schwarzschild-de Sitter metric \citep{ker+al03}.

Actually, variations in the orbital radius are pretty small. Over the life span of the solar system ($\sim 10^{17}$s) and assuming an Einstein-de Sitter universe with $h=0.5$, the fractional change of an Earth-like orbit ($ P \sim 1$~year) is of order of $10^{-23}$ \citep{coo+al98}. Considering a span of time suitable for observations, $\sim 10$~years, the orbital axis of a Pluto-like planet ($M \sim M_\odot$, $a \sim 30$~AU) changes by $\sim 2\times 10^{-14}$~m ($\sim 5\times 10^{-15}$~m) for an Einstein-de Sitter model (reference $\Lambda$CDM model) with $h=0.7$.  The accuracy required to detect such small changes is well above the reach of next generation technological facilities. The today best accuracy on the orbital semi-major axis in the Solar system, achieved for the inner planets, is of order of $10^{-1}$~m \citep{pit05}. Variations get bigger for phantom cosmologies, but not enough. For a flat models of universe with $\Om_\mathrm{M0} \ls 0.1$ and $w_\mathrm{X0} \ls -3$, changes are 50 times bigger, see Fig.~\ref{OmegaM_wX_var}.

Variations with time of the orbital radius are routinely observed in binary pulsars. As well known, orbital periods decrease in time due to the loss of energy by gravitational radiation emission. Cosmology effects produce another mechanism that can affect the orbital period variation,
\beq
\frac{\dot{P}}{P} =\frac{3}{2}\frac{\dot{a}}{a}.
\eeq 
Let us consider the prototype binary pulsar, B1913+16 ($P \sim 7.8$~hours). Its orbital period is observed to be decreasing with a rate of $\dot{P} = -2.4 \times 10^{-12}$ with an uncertainty of $\ls 10^{-15}$ \cite{we+ta05}. In a flat model of universe with $\Om_\mathrm{M0} \sim 0.3$ and phantom energy with $w_\mathrm{X0} = -2$, the orbital period variation is expected to be $\sim 10^{-40}h^3$, well below the experimental accuracy. The measurement of this effect in the near future seems very unlikely. 

The variation in the eccentricity can be expressed as
\beq
\left\langle \frac{d e}{dt} \right\rangle = \frac{1}{2a} \frac{\sqrt{1-e^2}}{e} \left\langle \frac{d a}{dt} \right\rangle \label{pla10} .
\eeq
An expanding orbit turns more eccentric with time, as expected from angular momentum conservation. As for the semi-major axis, changes are really small.

A more feasible approach to detect the effect of cosmological acceleration on planetary motion should be to investigate departures from the third Kepler's law. As can be seen from Eq.~(\ref{eul13}), due to the perturbing acceleration term the observed angular frequency will differ form the mean motion $n$. It is
\ba
\frac{\d n}{n} & = & - \frac{1}{2} \frac{1}{n^2} \frac{\ddot{R} }{R} \\
 & = &  \frac{1}{n^2} H^2 q ,
\ea
where we have introduced the deceleration parameter, $q$, in the second line. At present time
\beq
q_0 = \frac{1}{2} \left\{ \Om_\mathrm{M0}+(1+3 w_\mathrm{X0})\Om_\mathrm{X0} \right\} .
\eeq
The variation is now proportional to the acceleration of the scale factor. The statistical error on the mean motion for each major planet can be evaluated from the uncertainty on the semi-major axis, $\delta n = - (3/2) n \delta a/a$, and can then be translated into an uncertainty on the effective acceleration. When attempting to detect exotic physics in the Solar system with today data on changes in the mean motion, best bounds come from Earth and Mars \citep{se+je06a,se+je06b}. In our reference $\L$CDM model, the today deceleration parameter is $q_0 = -0.55$. Considering the Earth orbit ($\delta a \sim 0.1$~m, $a \sim 1$~AU), we get a limit $|q_0| \ls 0.8 \times 10^{10}/h^2$, ten orders of magnitude larger than estimates from observational cosmology. Constraints could greatly improve by considering a similar uncertainty on the measurement of the semi-major axis of outer planets, as could be obtained by dedicated radio ranging observations. For the Neptune orbit ($a \sim 30$~AU), an uncertainty $\delta a \sim 0.1$~m would imply  $| q_0| \ls 0.6 \times 10^{5}/h^2$, a steady improvement of nearly five orders of magnitude.

The effect of cosmology on the periastron precession is of the same order of the change in the mean motion. All cosmological models with a non null acceleration induce a variation in $\omega$, 
\beq
\left\langle \frac{d \omega}{dt} \right\rangle =\frac{3}{2 n}\sqrt{1-e^2}\frac{\ddot{R} }{R} .
\eeq
In fact, differently from axis radius and eccentricity, the main contribution comes from the factor $\ddot{R}/R$ and not from its variation in time. At present time, in terms of cosmological parameters,
\beq
\left\langle \frac{d \omega}{dt} \right\rangle_{z=0}= -\frac{3}{2 n}\sqrt{1-e^2} H_0^2 q_0
\eeq
In the case of a universe filled with only pressure-less dark matter with density $\rho_M$ and no dark energy, $\ddot{R}/R =-4\pi G \rho_M/3 $ and we retrieve the result in \cite{se+je06b}. In a de Sitter model ($\Om_\mathrm{M0}=0$), then $\ddot{R}/R =c^2 \L/3$, and we retrieve the well known result for extra-precession due to a cosmological constant as usually obtained in the framework of the Schwarzschild-de Sitter metric \citep{ker+al03,se+je06a}. 

The rate of precession predicted for the Earth is $\sim -5\times 10^{-16} q_0 h^2$~arcsec per year, $\sim 10^{-16}$~arcsec per year in our reference $\Lambda$CDM model with $h=0.7$. These estimates are ten orders of magnitude smaller than the state of art accuracy of $\sim 10^{-6}$~arcsec per year obtained for the Earth and Mars orbits \cite{pit05b}, so that the consequent upper bound on $q_0$ is of the same order of that derived from changes in the mean motion. We remark that dark matter and dark energy work in opposite directions. Constraints on the value of $\Lambda$ from perihelion precession and mean motion in the Solar system have been obtained so far considering the contribution of the cosmological constant to the field equations but neglecting the contribution from diffuse dark matter \citep[and references therein]{je+se06,se+je06a}. The contribution of diffuse dark matter should be accounted for. Furthermore, the Solar system is supposed to be embedded in a Galactic dark halo with a density in excess of nearly five orders of magnitude with respect to the mean cosmological dark matter. This local enhancement makes the measurement of dark matter with accurate planetary astrometric data at the reach of future experiments \cite{se+je06b}.

\section{Conclusions}
\label{sec:con}

Cosmological background can affect the orbit of a planetary system through the gravitational action of the smooth energy-matter components in which the system is embedded. Changes in the mean motion and anomalous periastron precession are sensitive to the acceleration of the scale factor, $\ddot{R}/R$, whereas variations in the semi-major axis and eccentricity come from its time variation. In principle, a very detailed knowledge of the orbital parameters of a bound gravitational system would allow to put constraints on the cosmological parameters through two diagnostics, i.e. $q H^2$ and $d (q H^2)/dz$, that are independent of the main ways in which observational cosmology usually investigate the energy content of the universe, i.e. via either the distance-redshift relation or the growth of structure. This can be appealing on a theoretical side with its promise in breaking degeneracies which affect the other methods, but it seems an unlikely tool if we consider the observational prospects. Actual bounds on the deceleration parameter from accurate astrometric data of perihelion precession and changes in the third Kepler's law in the Solar system fall short of ten orders of magnitude with respect to estimates from observational cosmology. Even if we consider other tests of gravity physics, such as geodetic precession and gravitational redshift, the situation should not improve substantially. In a previous paper \citep{se+je06a}, we discussed stellar system tests for the cosmological constant. Either future measurements of periastron advance in wide binary pulsars or gravitational redshift of white dwarfs could provide bounds competitive with Earth and Mars data but this is not enough to constrain cosmological parameters. Better prospects come from future radio-ranging measurements of Solar system outer planets which could improve actual bounds by five orders of magnitude. Such an improvement can be significant but still not competitive with observational cosmology.

Apart from the technological challenge required to reach the experimental accuracy needed to study cosmology in the Solar system, two other issues should be considered. In our analysis, we have considered a planetary system in an otherwise strictly homogeneous and isotropic universe. As a matter of fact, planetary systems are embedded in galactic dark matter haloes whose density is in excess of several orders of magnitude with respect to the mean cosmological dark matter. Furthermore, we have considered an homogeneous background but tidal forces from the environment could easily play a role. These two effects on the galactic scale should easily overcome the effects of the acceleration of the cosmological scale factor. Any future observed deviation in the measured values of perihelion precession or in the third Kepler's law should then be seen as an indication for Galactic dark matter rather than an effect of an accelerated expanding universe. 

Till now we have considered effects due to the expansion only on planetary orbital motions. It might be interesting to consider also possible space-based interferometric type experiments similar to the one proposed to detect gravitational waves as is the case of the Laser Interferometer Space Antenna (LISA) \footnote{http://www.lisascience.org}. The radius of an Earth-like orbit would increase, assuming our reference $\Lambda$CDM universe with $h=0.7$, by $\sim 6 \times 10^{-22}$~m per year. Accordingly, for a spacecraft configuration like the one envisaged for LISA, which orbits around the Sun at the same distance as the Earth and with a separation of $\sim 5 \times 10^6$~km between two modules, it would imply an increase of the separation of the order of $\sim 2 \times 10^{-23}$~m per year. In LISA a laser beam sent by a module bounces back from the other module's free falling mirror and then gets back to the original module where, after having bounced off the mirror, the light is mixed with a fraction of the outgoing light, and then interference is detected. In principle, an increase in the separation distance between two modules could be detected in this way but would require that the laser beam maintains a very high phase stability during years, which is well beyond today technological feasibility.

\begin{acknowledgments}
M.S. is supported by the Swiss National Science Foundation and by the Tomalla Foundation.
\end{acknowledgments}


\end{document}